\documentclass[conference]{IEEEtran}
\IEEEoverridecommandlockouts
\usepackage{cite}
\usepackage{amsmath,amssymb,amsfonts}
\usepackage{algorithmic}
\usepackage{graphicx}
\usepackage{textcomp}
\usepackage{xcolor}
\usepackage{mdframed}
\def\BibTeX{{\rm B\kern-.05em{\sc i\kern-.025em b}\kern-.08em
    T\kern-.1667em\lower.7ex\hbox{E}\kern-.125emX}}

\usepackage[absolute]{textpos}
\usepackage{hyperref}
\begin{document}

\IEEEoverridecommandlockouts
\title{ {\fontsize{13}{13}\selectfont \textbf{This paper was presented at IEEE VLSI Test Symposium (VTS) 2025}} \\[0.5em] 
BugWhisperer: Fine-Tuning LLMs for SoC Hardware Vulnerability Detection\\
\thanks{We thank to U.S. National Science Foundation (NSF) for their support through CAREER
Award under Grant 2339971.}

\author{\IEEEauthorblockN{Shams Tarek, Dipayan Saha, Sujan Kumar Saha, Farimah Farahmandi}
\IEEEauthorblockA{\textit{Department of Electrical and Computer Engineering, University of Florida, Gainesville, FL, USA}\\
\{shams.tarek, dsaha, sujansaha\}@ufl.edu, \{farimah\}@ece.ufl.edu}}}

\maketitle

\begin{abstract}
The current landscape of system-on-chips (SoCs) security verification faces challenges due to manual, labor-intensive, and inflexible methodologies. These issues limit the scalability and effectiveness of security protocols, making bug detection at the Register-Transfer Level (RTL) difficult. This paper proposes a new framework named BugWhisperer that utilizes a specialized, fine-tuned Large Language Model (LLM) to address these challenges. By enhancing the LLM's hardware security knowledge and leveraging its capabilities for text inference and knowledge transfer, this approach automates and improves the adaptability and reusability of the verification process. We introduce an open-source, fine-tuned LLM specifically designed for detecting security vulnerabilities in SoC designs. Our findings demonstrate that this tailored LLM effectively enhances the efficiency and flexibility of the security verification process. Additionally, we introduce a comprehensive hardware vulnerability database that supports this work and will further assist the research community in enhancing the security verification process.

\end{abstract}


\section{Introduction}
Hardware vulnerabilities at the SoC level present critical threats by potentially exposing sensitive user data, cryptographic keys, and essential system configurations. Increasingly sophisticated attacks targeting SoCs, such as information leakage \cite{contreras2017security}, side-channel leakage \cite{mishra2017security}, and violations of access control \cite{property}, emphasize the urgent need for rigorous SoC security verification. Addressing these vulnerabilities proactively is essential to prevent severe financial repercussions, product recalls, and reputational damage within the semiconductor industry. 

Traditional SoC security verification approaches primarily concentrate on functional correctness, frequently overlooking critical vulnerabilities during pre-silicon verification. While methods such as information flow tracking \cite{ift}, assertion-based security verification \cite{property}, fuzz testing \cite{fuzz}, runtime verification monitoring, and static code analysis \cite{arc_fsm_g} have been developed, these techniques often encounter significant scalability and adaptability limitations. Furthermore, they usually require extensive manual intervention, thereby elevating both the complexity and cost of security verification. Consequently, there is an increasing demand for automated and versatile methods capable of efficiently handling diverse hardware architectures and dynamic security scenarios. Detecting vulnerabilities comprehensively at the RTL can notably reduce the time, effort, and expenses involved in SoC security verification.

Recently, LLMs have gained traction within hardware security and design communities due to their superior abilities in pattern recognition, knowledge generalization, and learning from extensive datasets \cite{socurellm,access_paper}. Consequently, researchers are leveraging LLMs to address complex security challenges at the SoC level \cite{ChipGPT,access_paper,jv, hadi}. For instance, prompting-based techniques utilizing both pre-trained and proprietary LLMs have been applied to identify security vulnerabilities within RTL designs \cite{host_paper}. Nevertheless, the limited hardware-specific knowledge inherent in pre-trained LLMs constrains their applicability in targeted security verification tasks.
\begin{figure*} [!t] 
    \centering
    \includegraphics [scale=.7]
    {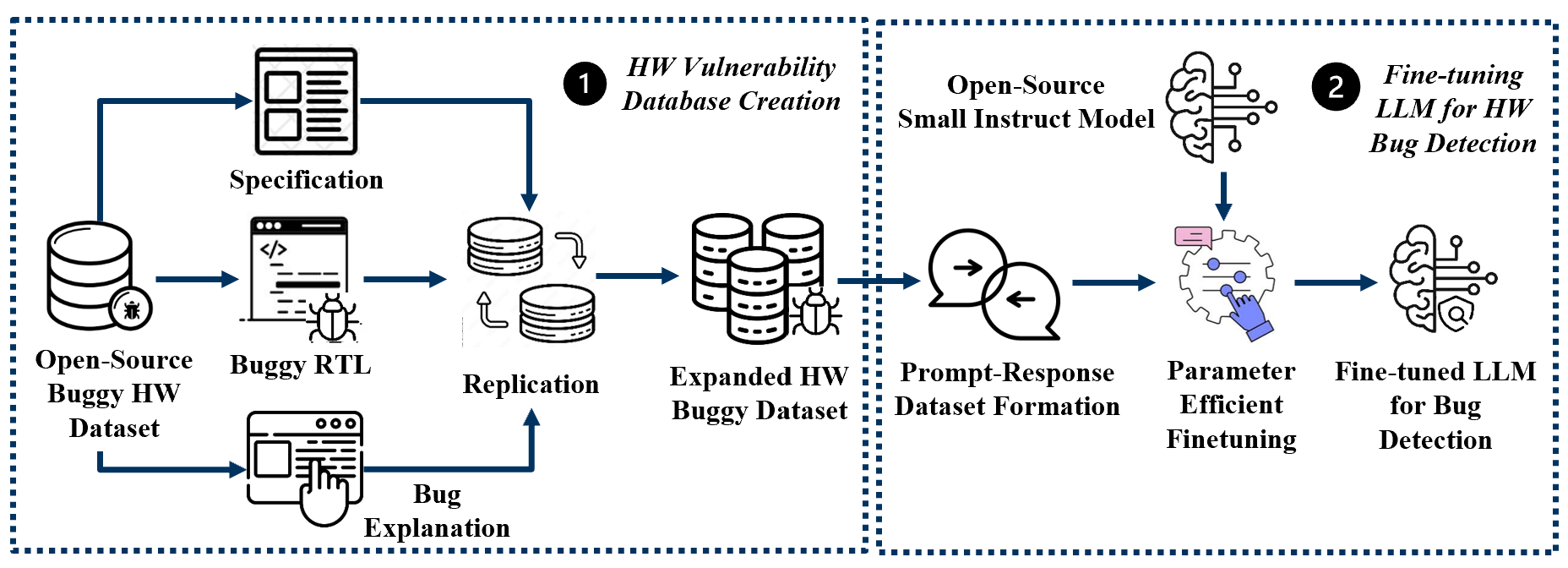} 
    \caption{Overview of the proposed BugWhisperer framework.}
    \label{fig1}    
\end{figure*}
A comparative analysis of existing LLMs provides valuable insights into their effectiveness for vulnerability detection SoCs and hardware designs. Proprietary models, including well-known solutions such as Gemini and GPT \cite{gpt4}, typically demonstrate superior detection accuracy compared to smaller open-source alternatives like LLama \cite{llama3}, Mistral, and CodeLlama \cite{codellama}. However, despite their notable performance advantages, proprietary solutions present several significant challenges. These include high operational costs, limited accessibility, scalability constraints, and reduced operational flexibility. Such factors may substantially limit their broader adoption and applicability across various industries, potentially making open-source alternatives more attractive due to their flexibility, transparency, and accessibility.

Thus, the development of smaller, open-source LLMs explicitly fine-tuned to detect SoC-level security vulnerabilities in RTL designs becomes critical. To our knowledge, no such model currently exists. This paper addresses this significant gap by investigating the feasibility and effectiveness of fine-tuning an open-source LLM for hardware security vulnerability detection at the SoC level. By integrating comprehensive insights from the CWE hardware vulnerability database, our proposed model substantially enhances its capability to detect and mitigate critical security threats.

In addition, the model is trained using security vulnerability reports, which allows it to accurately identify the presence of security vulnerabilities. Using these task-specific datasets, our fine-tuned model improves the precision and efficiency in detecting 13 security vulnerabilities in SoC designs. 
The key contributions are: 
\begin{itemize}
    \item We present a comprehensive SoC hardware vulnerability database, open to the research community
    \item We introduce an open-source, fine-tuned LLM specifically designed for detecting hardware design bugs in the SoC
    \item We demonstrate that open-source LLMs can compete with proprietary LLMs if properly curated
\end{itemize}

In the remainder of this paper, Section \ref{sec:methodology} narrates the methodology used in this work. Next, Section \ref{sec:experiment} describes the training scheme and details the experimental result with analysis. Finally, Section \ref{sec:conclusion} concludes the paper.

\section{Proposed Methodology} \label{sec:methodology}

The proposed framework ``BugWhisperer" comprises two key stages: (1) Generation of a hardware vulnerability database, and (2) Fine-tuning of a custom LLM for hardware vulnerability detection, as shown in Figure \ref{fig1}.

\subsection{Database Generation for Vulnerable SoC Designs}

To establish a high-quality database for vulnerable hardware designs, this work leverages existing Golden Vulnerable SoC design benchmarks available through the Cad4Security platform \cite{benchmark}. These benchmarks comprise 13 distinct SoC vulnerabilities introduced into the CVA6 Ariane core. Initially, each vulnerable design module is systematically separated and clearly labeled according to its specific vulnerability type. Subsequently, a comprehensive specification file is generated for each vulnerable module. Each specification file meticulously documents the baseline functionality of the respective Intellectual Property (IP), detailed descriptions of internal and external registers, Input/Output (I/O) ports, and explicit vulnerability characteristics. Such specification documentation is crucial, as it explicitly outlines both the intended baseline functionality and the inherent vulnerabilities within each module.

Following specification creation, the vulnerable designs are replicated using a Replicator LLM. This replication is not a simple duplication; rather, it involves a nuanced process engineered to retain original module functionality while systematically diversifying code expression. To achieve this diversity, various Verilog/SystemVerilog coding styles are adopted, including parameterization, Finite State Machine (FSM) architectures (single-process FSM and dual-process FSM), and varying signal nomenclatures. The replication process employs carefully curated Replicator prompts, manually tailored according to these different coding styles, to ensure each generated design instance remains distinct.

During replication, the specification file corresponding to each vulnerable IP is provided as context to the Replicator LLM, thereby ensuring fidelity to the original functionality and the embedded vulnerability. Furthermore, two pivotal LLM parameters—``Temperature" and ``Top\_p"—are utilized to enhance the uniqueness and diversity of the generated design instances. The ``Temperature" parameter, adjustable between 0 and 2, influences response creativity and diversity, with typical optimal ranges identified as between 0.6 and 1.5. Higher temperature values increase diversity but may compromise functionality consistency due to enhanced creativity. The ``Top\_p" parameter controls randomness and coherence in the output tokens, aiding in the prevention of code similarity and token overlap among generated replicas.

\subsection{Fine-tuning LLMs for Vulnerability Detection}

In the subsequent stage, open-source LLMs are fine-tuned for vulnerability detection tasks using the developed hardware vulnerability database. Initially, potential open-source models are carefully analyzed for suitability. Many open-source LLMs lack comprehensive hardware-specific domain knowledge essential for the effective detection of vulnerabilities in hardware designs. After rigorous assessment, we select \textit{Llama-3.2-1B-instruct}, \textit{Llama-3.2-3B-instruct}, \textit{Llama-3.1-8B-instruct}, \textit{Mistral-7B-instruct}, and \textit{Codellama-7B-it} models. These models are chosen due to their instructional adaptability, coding proficiency, and compatibility with hardware design tasks.

Despite their robust coding and instructional following capabilities, these models initially lack specialized knowledge regarding hardware security vulnerabilities, especially in RTL design contexts. Therefore, fine-tuning is necessary to embed this critical domain-specific knowledge. The fine-tuning process enhances the LLMs' capability to accurately recognize and classify hardware security vulnerabilities, thereby significantly improving their efficiency and reliability during security bug analysis. The incorporation of domain-specific expertise enables these models to effectively support hardware security assessments, contributing substantially to the robustness and security of future hardware designs. The detailed description of the vulnerabilities will be found here \cite{benchmark}. The fine-
tuning and evaluation efforts discussed in this work focus on detecting the following
13 vulnerabilities: 

\begin{itemize}
    \item CWE-1198: Improper handling of privilege issues
    \item CWE-269: Improper privilege level during interrupt handling
    \item CWE-1245: Less secured FSM encoding
    \item CWE-1260: Overlapping between memory ranges
    \item CWE-506: Hardware trojan inside the decoder module
    \item CWE-310: Trojan in AES for information leakage
    \item  CWE-310: Trojan in AES for denial of service
    \item CWE-310: Trojan in CSR module unauthorized access
    \item CWE-321: Use of hardcoded cryptographic key 
    \item CWE-250: Improper trap privilege assignment
    \item CWE-1244: Unlocking JTAG during reset 
    \item CWE-284: Improper direct memory access
    \item CWE-1271: Unauthorized access to important registers

\end{itemize}
The detailed description of the vulnerabilities will be found here \cite{benchmark}.

During this phase, instruction fine-tuning is utilized to specialize the model for identifying security vulnerabilities. To accomplish this, the original data is reformatted via a Python script into a structured prompt-response format. Each prompt includes either a secure or a vulnerable SoC design accompanied by a query prompting a security evaluation targeting a particular security flaw. In the corresponding responses, it is clearly stated whether the specified vulnerability is present or not, along with explanations to support the decision. Such detailed rationales are essential during fine-tuning, as they assist the model in comprehending the logic underlying the security assessments, thus significantly improving its capability to identify and clearly articulate security issues within SoC designs.

A major challenge in dataset creation is the lack of sufficiently detailed annotations. To mitigate this limitation, GPT-4o is employed to generate thorough explanatory annotations across different hardware modules. These generated explanations are subsequently integrated into the prompt-response pairs, thereby enhancing the dataset with the necessary context for effective model training. For a rigorous evaluation of the model’s effectiveness, the dataset is partitioned into distinct training, validation, and testing subsets based on individual designs. Such an approach ensures that the RTL designs allocated to the test set remain entirely unseen during the training phase, thereby enabling an accurate assessment of the model's ability to generalize beyond the provided training examples.

\begin{figure*} [h] 
    \centering
    \includegraphics [scale=.56]
    {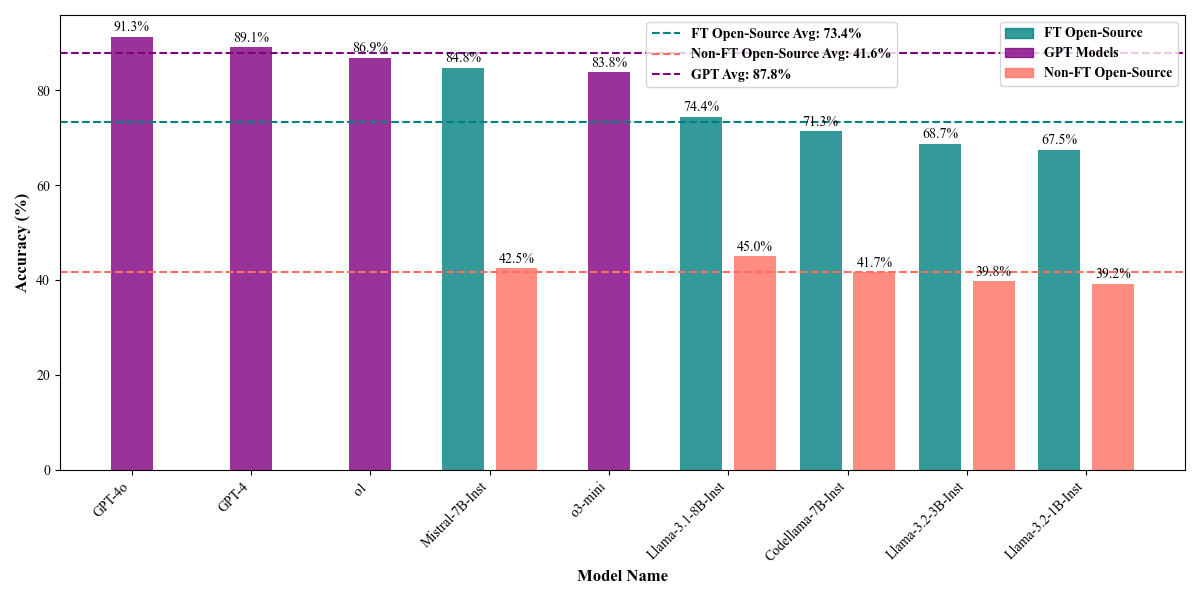} 
    \caption{Performance evaluation of the large proprietary models, fine-tuned and non-fine-tuned open-source models.}
    \label{data}    
\end{figure*}

\section{Experimentation and Evaluation} \label{sec:experiment}
\subsection{Dataset Replication}
\label{sec:dataset_replication}
For replicating the dataset, available open-source SoC vulnerability benchmarks were used. The coding and replicating capabilities of GPT-4 and GPT-4o were leveraged as the replicator LLMs. All datasets were replicated using GPT API calls. To maintain consistent functionality throughout the process, the ``Temperature" value was kept between 0.6 and 1.5. Using this process, a set of 4000 vulnerable SoC hardware designs were generated. The dataset is available here: \url{https://github.com/shamstarekargho/Hardware-Vulnerability-Dataset}

\subsection{Training Setup}
As described in Section \ref{sec:dataset_replication}, 4,000 Verilog codes are used during training the open-source models. For training, We implemented a parameter-efficient fine-tuning approach using low-rank adaptation (LoRA) \cite{lora}, enhancing training efficiency by adding minimal additional parameters to the open-source models. The LoRA configuration was set with a rank size of 128, an alpha value of 256, and a dropout rate of 0.1, optimizing the balance between parameter efficiency and model accuracy. Fine-tuning was conducted on two NVIDIA A100 GPUs with 4-bit quantization (NF4) and a float16 compute data type to manage memory demands, leveraging the GPUs’ computational power. To ensure stable weight updates, we employed a low learning rate of $2 \times 10^{-6}$, a batch size of 4, and gradient accumulation steps of 1 across three training epochs, mitigating overfitting risks while maintaining effective gradient updates. The training utilized the paged adamw 32bit optimizer with a weight decay of 0.001 for regularization, a maximum gradient norm of 0.3 for clipping to prevent exploding gradients, and a constant learning rate scheduler with a warmup ratio of 0.03 to gradually increase the learning rate. Gradient checkpointing was enabled to reduce memory usage, and the maximum sequence length was capped at 512 tokens to balance computational efficiency and context retention.
\subsection{Result Analysis}
The detection accuracy results for different models, including proprietary, fine-tuned, and non-fine-tuned open-source models, are shown in Figure \ref{data} The results indicate a clear distinction in performance between these categories, demonstrating the effectiveness of fine-tuning in adapting LLMs for hardware security vulnerability detection.

Among the fine-tuned models, \textit{Mistral-7B-instruct} achieved the highest accuracy of 84.8\%, significantly outperforming its non-fine-tuned counterpart, which only reached 42.5\%. This improvement of over 40 percentage points highlights the effectiveness of fine-tuning in equipping open-source models with domain-specific knowledge. Although \textit{Mistral-7B-instruct} does not surpass the performance of proprietary models, its high accuracy suggests that open-source models can be viable alternatives with sufficient domain adaptation. Similar trends are observed for \textit{Llama-3.1-8B} and \textit{Llama-3.2-3B}, where fine-tuning elevates detection accuracy to 74.4\% and 68.7\%, respectively, from their significantly lower non-fine-tuned baselines. The best-performing model (Mistral-7b-instruct-Bug-Whisperer) has been released to the research community for further use in the hardware security verification domain. The model can be found here: \url{https://huggingface.co/shamstarek/Mistral-7B-instruct-Bug-Whisperer}

The proprietary models, as expected, demonstrated superior performance, with \textit{GPT-4o} achieving the highest accuracy of 91.3\%, followed by \textit{o-1} (86.9\%) and \textit{o3-mini} (83.8\%). These results align with expectations, as proprietary models benefit from extensive pretraining on diverse datasets, including specialized knowledge in hardware security. However, their reliance on closed-source architectures and significant usage costs make them less accessible for widespread deployment.

A key observation from the results is the poor performance of non-fine-tuned open-source models, which average around 40\% accuracy. This highlights a major limitation of general-purpose LLMs when applied to domain-specific tasks without adaptation. The lack of pretraining on hardware security datasets renders them ineffective for vulnerability detection, reinforcing the necessity of fine-tuning to bridge this knowledge gap.

Model size is another influential factor affecting detection accuracy. The results indicate that larger open-source models tend to achieve higher accuracy post-fine-tuning. For instance, \textit{Llama-3.2-3B} (3B parameters) reaches 68.7\% accuracy, whereas \textit{Llama-3.1-8B} (8B parameters) achieves 74.4\%, suggesting that increased model capacity contributes to better representation of security-related patterns. However, fine-tuning smaller models effectively can still yield competitive performance while reducing computational overhead.

These findings suggest broader implications for AI-driven hardware security. The substantial improvement in fine-tuned models highlights their potential as cost-effective alternatives to proprietary solutions. Unlike closed-source models, fine-tuned open-source LLMs provide transparency, cost efficiency, and flexibility for integration into security workflows.
However, Despite these promising improvements, limitations remain. While fine-tuning enhances detection accuracy, some vulnerabilities may still go undetected due to complex attack patterns. Further refinements in dataset expansion, fine-tuning methodologies, and architectural modifications could push open-source models closer to proprietary-level performance.

\section{Conclusion} \label{sec:conclusion}
This paper presents a fine-tuned LLM-based approach, ``BugWhisperer", to SoC security verification, addressing the limitations of manual and inflexible methodologies. The proposed model significantly improves vulnerability detection accuracy, outperforming non-fine-tuned counterparts by over 40\% and demonstrating the potential of open-source LLMs as cost-effective alternatives to proprietary solutions. Additionally, the introduction of a comprehensive hardware vulnerability database enhances research in automated security verification. These findings highlight the scalability and adaptability of LLMs for hardware security, with future work focusing on further optimization and dataset expansion.

\bibliographystyle{IEEEtran}
\bibliography{IEEEabrv,bugwhisperer}

\end{document}